\documentclass[sigconf, nonacm]{acmart}
\usepackage{tabularx}
\usepackage{colortbl}

\usepackage{graphicx} % Required for inserting images
\usepackage{threeparttable}
\usepackage{appendix}
\usepackage{subcaption}
\usepackage{hyperref} % Add this to the preamble if not already included

\renewcommand\footnotetextcopyrightpermission[1]{} % Removes copyright notice

\makeatletter
\def\NAT@spacechar{~}% NEW
\makeatother

\title{NetAurHPD: Network Auralization Hyperlink Prediction Model to Identify Metabolic Pathways from Metabolomics Data}
\author{Tamir Bar-Tov, Rami Puzis, David Toubiana}
\date{August 2024}

\begin{abstract}
Metabolite biosynthesis is regulated via metabolic pathways, which can be activated and deactivated within organisms. Understanding and identifying an organism’s metabolic pathway network is a crucial aspect for various research fields, including crop and life stock breeding, pharmacology, and medicine. The problem of identifying whether a pathway is part of a studied metabolic system is commonly framed as a hyperlink prediction problem. The most important challenge in prediction of metabolic pathways is the sparsity of the labeled data. This challenge can partially be mitigated using metabolite correlation networks which are affected by all active pathways including those that were not confirmed yet in laboratory experiments. Unfortunately, extracting properties that can confirm or refute existence of a metabolic pathway in a particular organism is not a trivial task. In this research, we introduce the Network Auralization Hyperlink Prediction (\textbf{NetAurHPD}) – a framework that relies on (1) graph auralization to extract and aggregate representations of nodes in metabolite correlation networks and (2) data augmentation method that generates metabolite correlation networks given a subset of chemical reactions defined as hyperlinks. Experiments with metabolites correlation-based networks of tomato pericarp demonstrate promising results for NetAurHPD, compared to alternative methods.  Furthermore, the application of data augmentation improved NetAurHPD's learning capabilities and overall performance. Additionally, NetAurHPD outperformed state-of-the-art method in experiments under challenging conditions, and has the potential to be a valuable tool for exploring organisms with limited existing knowledge.
\end{abstract}

\begin{document}

\maketitle

\section{Introduction}
Understanding metabolic behavior is essential across various life science fields, including crop breeding improvement, drug design, and bio-fuels \cite{jang2022applications,ursem2008correlation,zhang2016recovering,bordbar2014constraint}. 
Metabolism describes the complex biochemical processes occurring within a cell, which generally convert materials entering the cell into substances necessary for cellular and organismal growth and function \cite{dale2010machine,aljarbou2020determining}.

Metabolic pathways are described as linked series of chemical reactions, where each reaction involves the conversion of one or more molecules, known as substrates, into one or more molecules, known as products. 
In the context of metabolism these molecules are called metabolites \cite{toubiana2013network}. 
While many metabolic pathways have been well-characterized, there is still much to discover \cite{monk2014optimizing}.

% The field of metabolism research has undergone significant advancements. In the past century, research in genetics, biochemistry, and molecular biology commonly focused on studying individual components of biological systems, often neglecting the interactions among these components and the biological system. 
% However, with the advent of new technological capabilities and the reduction of data collecting costs, researchers now have access to vast datasets that were previously unavailable. 
% To cope with the increasing volume of multidimensional data, novel methods have been developedand enabled researchers to construct high-throughput models capable of simultaneously targeting multiple aspects of metabolism \cite{bordbar2014constraint}.

The reduction of data-collecting costs and the advent of new data-mining technologies facilitate the construction models targeting multiple aspects of metabolism~\cite{bordbar2014constraint}.  
In this study, we focus on two types of data:  (1) the metabolic network and (2) the metabolite correlation networks. 
Metabolic networks, hypergraphs where nodes are metabolites and directed hyperlinks represent chemical reactions,
capture the known metabolic pathways of a particular organism  \cite{toubiana2013network,aljarbou2020determining}. 
Hyperlink prediction methods can be applied to metabolic networks to identify hypotheses of new unknown metabolic pathways \cite{chen2023Teasing}. 
However, metabolic networks cannot indicate whether or not the chemical reactions within a pathway took place in a set of samples.         

Metabolite correlation networks (\textbf{Metabolite CN})  capture the correlations between the quantities of metabolites \cite{toubiana2019combined}. 
Metabolite CNs are constructed based on a high throughput collection of metabolic profiles of the organisms and do not depend on prior knowledge of the metabolic pathways.   
Contrary to the metabolic networks, Metabolite CNs are believed to provide a ‘fingerprint’ of an organism's underlying biochemical system and its regulatory mechanisms reflecting chemical reactions that actually happened within a set of samples~\cite{ursem2008correlation}. 
This assumption is the theoretical justification of the current study. Nevertheless, extracting information about chemical reactions from  Metabolite CNs is a major challenge due to the extreme complexity of the problem~\cite{steuer2003observing}.

This research aims to exploit the information embedded in the Metabolite CN to predict the presence and absence of metabolic pathways in an organism. 
%Specifically, we aim to develop a hyperlink predictor that relies on the pathways classifier capable of predicting the presence and absence of metabolic pathways based on Metabolite CN constructed from tomato pericarp samples. 
Specifically, we develop the network auralization hyperlink prediction model (\textbf{NetAurHPD}), which relies on a waveform representation of metabolites within the Metabolite CN, combined with deep learning (\textbf{DL}) methods. 
The waveform representation is obtained through the process of Network Auralization~\cite{li2023centrality}. 
The models are evaluated on Metabolite CNs of tomato pericarp ~\cite{toubiana2019combined}.
To mitigate the challenge of data sparsity, we suggest a simple data augmentation process that generates artificial Metabolite CNs given a subset of known reactions.   
Our work provides the following contributions:
\begin{enumerate}
  \item We present NetAurHPD, a new hyperlink prediction method based on auralized Metabolite CN, which demonstrated promising results.
  \item Given that many organisms have a limited number of known pathways, we propose a data augmentation method to enrich the training data, and improve both the performance and robustness of NetAurHPD. 
  \item We demonstrate the advantage of NetAurHPD over state-of-the-art hyperlink prediction methods in challenging real-world biological research conditions.
\end{enumerate}

% The second goal of this work is to develop a data augmentation method to enlarge the training dataset. 
% For many organisms, the amount of known pathways is small. The need for a data augmentation method is evident, especially when dealing with DL models, which typically require large datasets to perform well. 
% In this research, we will augment the dataset by imitating the organism’s metabolic activity and generate Metabolite CNs (\textbf{Augmentations}).

% Lastly, we will demonstrate the significance of using Metabolite CN as an input. We believe that biologists exploring organisms with limited prior knowledge of known pathways, should use NetAurHPD rather than alternative methods that rely solely on pathway information.

\section{Background}
\label{chapter:Background}
This work integrates knowledge from various disciplines, including biology, graph theory, DL, and signal processing. In this section, we will discuss the relevant foundational concepts for our proposed research solution. 
\subsection{Hypergraphs}
\subsubsection{The Concepts of Hypergraphs}
Graphs are common structures for representing and analyzing relationships between entities. In countless real-world networks or systems, we can symbolize objects as nodes (or vertices), and any kind of relationship between two objects as a link (or edge). The links can be directed or undirected, depending on the symmetry of the relationship. However, some relationships involve more than two nodes. In such cases we can represent the relationship as a hypergraph - a graph in which a link can join together more than two nodes. In other words, a hyperlink (or hyperedge) is a subset of nodes. Naturally, a hypergraph can also be directed when an asymmetric relationship exists within the hyperlink participants \cite{zhou2006learning}. Mathematically, an unweighted hypergraph $H = \{V,F\}$, where $V = \{v_{\scriptstyle 1}, v_{\scriptstyle 2}, \ldots, v_{\scriptstyle n}\}$ stands for the node set and $F = \{f_{\scriptstyle 1}, f_{\scriptstyle 2}, \ldots, f_{\scriptstyle m}\}$ is the hyperlink set, and $F \subseteq \wp(V)$.

\begin{figure}[h]
    \centering
    \includegraphics[width=0.6\linewidth]{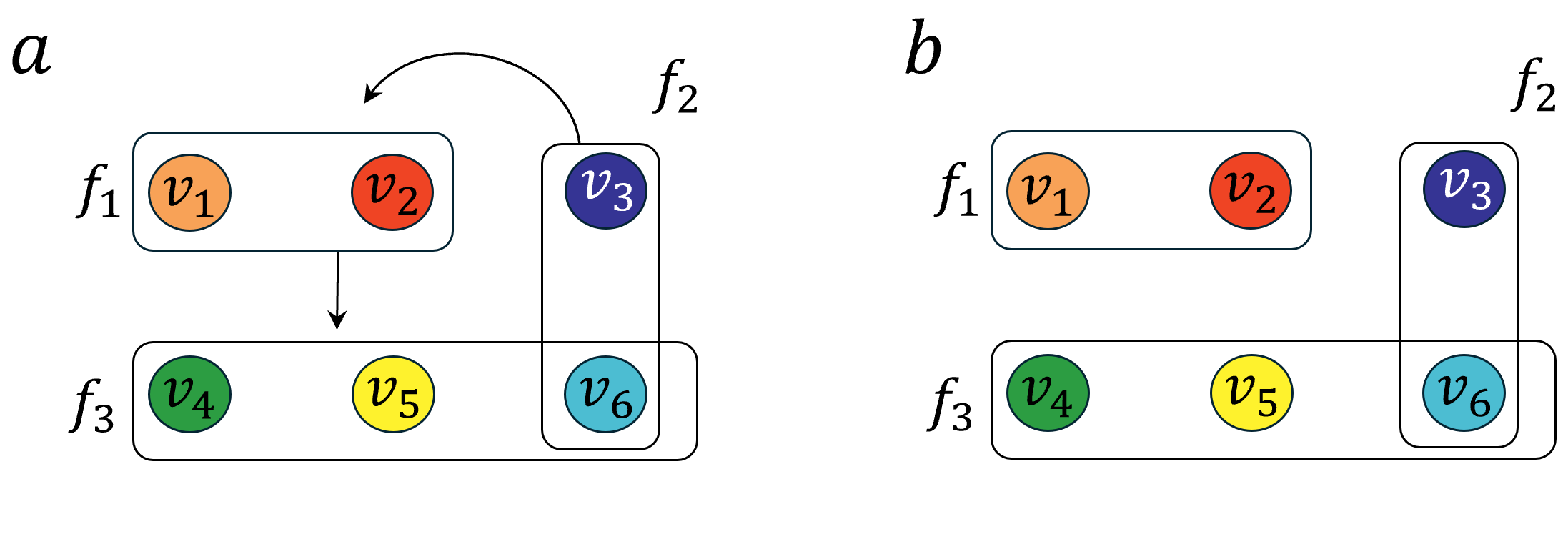}
    \caption{(a) Directed and (b) Undirected Hypergraph Structures}
    \label{fig:hypergraphs}
\end{figure}

\subsubsection{Hyperlink Prediction}
Link prediction is a well-explored researched challenge aimed at predicting absent connections in an incomplete graph or links with high potential to form, based on the observed links \cite{belth2022hidden}. Hyperlink prediction extends the link prediction task to hypergraphs \cite{chen2023survey}. In recent years we have seen extensive research dealing with this problem, a natural development considering the many real-world cases that can be modeled as hypergraphs. Formally, for a given hyperlink $f$ we aim to learn function $\Psi$ and threshold $\Theta$ such that:

\[
\Psi(f) =
\begin{cases} 
\geq \Theta & \text{if } f \in F, \\
< \Theta & \text{if } f \notin F.
\end{cases}
\]

As mentioned, a simple link assumes a fixed parameter of two nodes in each link while a hyperlink holds this parameter of nodes as a variable. This leads to a dramatic difference in the total number of hyperlink combinations, exponentially growing from $M^2$ combinations in simple graphs to $2^M$ in hypergraphs, where $M$ is the number of nodes. Consequently, covering all hyperlink candidates in a prediction task becomes computationally infeasible.

\subsubsection{Negative Sampling}
Most supervised link prediction models suffer from an imbalanced training set due to the sparsity of most real networks (few links vs many non-links). 
This problem is exaggerated in hypergraphs where non-links include arbitrary subsets of nodes. 
The common approach to mitigate this risk is negative sampling — undersampling for the absence of hyperlinks. We consider \( H_{\text{obs}} \) as the subset of observed hyperlinks, which is the given hypergraph, and denote the subset of unobserved hyperlinks as \( F_{\text{unobs}} \). Using negative sampling methods, we sample non-hyperlink examples \( \hat{F}_{\text{sam}} \subseteq \hat{F}_{\text{all}} := \wp(V) \setminus F \) \cite{patil2020negative}.

The selection of negative examples is crucial, as certain hyperlinks may be too straightforward for the model to predict, thereby potentially resulting in a weak model. Most negative sampling methods focus on generating candidates that resemble the structural parameters of positive hyperlinks, thereby forcing the model to better discern between classes and learn the nuances of the problem \cite{patil2020negative}. An optional method involves generating a corresponding negative hyperlink $f$ for each positive hyperlink $e$. Here, $\alpha$ is defined within the range of [0, 1], where $\alpha$×100\% of the nodes in $f$ are genuinely from $e$, while the rest are randomly sampled from the remaining nodes. The parameter $\alpha$ governs the authenticity of the negative hyperlinks; higher values of $\alpha$ signify that the negative hyperlinks are closer to the true ones \cite{yadati2020nhp}.

% \subsubsection{Hypergraph Transformations}
% Modeling with hypergraphs presents a unique challenge. The data structure often requires modifications, which can force a trade-off between simplifying assumptions and potential information loss. As described in Chen and Liu survey \cite{chen2023survey}, Several approaches exist to tackle this challenge. A very common method, known as Clique expansion, replaces each hyperlink with a clique, in which all nodes are directly linked to each other. Another approach involves converting the hypergraph into a bipartite graph, also
% known as Star expansion. This creates two sets of nodes: one representing the original nodes and another representing the hyperlinks. links are then drawn between corresponding nodes to hyperlink-nodes, indicating participation in a hyperlink. Line graph is another alternative, involving the creation of a new graph based on the original hypergraph, where every node represents a hyperlink.
% %\textcolor{red}{consider remove graph}
% \vspace{-7pt}
% \begin{figure}[h]
%     \centering
%     \includegraphics[width=0.6\linewidth]{images/approaches in hypergraph transformations.png}
%     \caption{Approaches in hypergraph expansions: (a) Original hypergraph (b) Clique expansion (c) Bipartite graph, also known as Star expansion (d) Line graph }
%     \label{fig:transformations}
% \end{figure}
% \vspace{-15pt}

\subsection{Metabolites Correlation Networks}
\label{section:Metabolites CN}
Metabolic profiling aims to quantify metabolites in a biological sample \cite{steuer2003observing}. Based on the metabolite profiles we construct a Metabolite CN, which is represented as a weighted graph \( G = (V, E, w) \). Here, \( V \) is the set of nodes representing the sampled metabolites in the biological sample, and \( E \) denotes the links between them. The links are derived from the metabolite’s pairwise similarity calculation of the metabolite profiles \cite{toubiana2013network}, often using the \textit{Pearson Correlation Coefficient}:
\vspace{-1pt}
\[
r(X, Y) = \frac{\sum_{i=1}^{n} (X_i - \bar{X}) (Y_i - \bar{Y})}{\sqrt{\sum_{i=1}^{n} (X_i - \bar{X})^2 \sum_{i=1}^{n} (Y_i - \bar{Y})^2}}
\]

An essential step in constructing the CN involves removing the spurious relationships between metabolites. One approach is to set thresholds \( \alpha \) and \( \beta \), where \( -1 < \alpha < 1 \) and \( 0 < \beta < 1 \), ensuring that only correlations that meet the conditions \( |r|\geq\alpha \) and $p-value$ \( < \beta \) are represented as links in \( G \). Following that, \( w \) corresponds to the selected similarity method which derives the link’s weight, in our example $r$. Alternatively, Metabolite CN can be constructed as an unweighted graph \( G = (V, E) \) \cite{toubiana2019combined}.

\begin{figure}[t]
    \centering
    \includegraphics[width=0.4\linewidth]{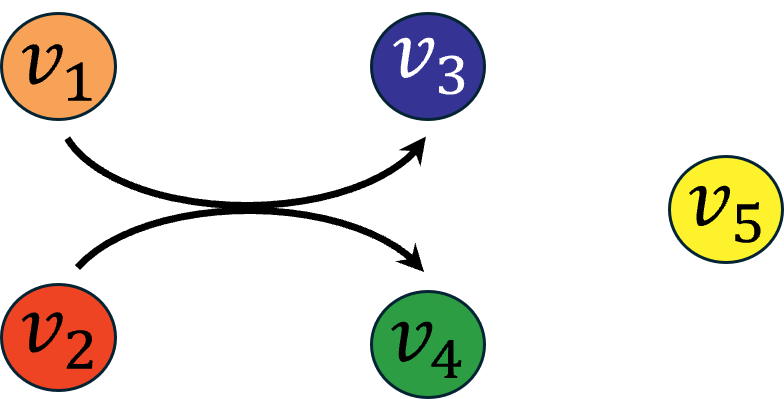}
    \caption{Example of one-step pathway}
    \label{fig:onestep}
\end{figure}

Theoretically, links’ weights are expected to correspond to the metabolite's roles as substrates and products. For example, assuming a one-step pathway as illustrated in Fig\ref{fig:onestep}, the substrates $v_{\scriptstyle 1}$ and $v_{\scriptstyle 2}$ appear together as part of the pathway substrates and therefore we expect a strong positive correlation between them. The same is expected for $v_{\scriptstyle 3}$ and $v_{\scriptstyle 4}$ as products. Given that $v_{\scriptstyle 1}$ and $v_{\scriptstyle 2}$ transform into $v_{\scriptstyle 3}$ and $v_{\scriptstyle 4}$, a decrease in the concentration of substrates should increase the concentration of products, leading to an expected strong negative correlation between the nodes on opposite sides of the pathway. Conversely, $v_{\scriptstyle 5}$ is expected to demonstrate weak correlations with other nodes, and its connections should not be regarded as links in $G$.

The described behavior is applicable within a "closed world" scenario, where only the specified pathway is considered. However, as the number of pathways increases, the simultaneous activity is challenging for the straightforward analysis outlined above. For example, a strong positive correlation between metabolites can occur even without a shared pathway, simply because they were active simultaneously. Also in metabolite level, participating in a few pathways simultaneously can lead to misleading similarity or dissimilarity. Considering the noise in real-world data, we better understand the challenges of analyzing Metabolite CNs in biological samples with hundreds of pathways, some of which may remain undiscovered.

Metabolomics is highly sensitive to changes in the environmental conditions of an organism that can lead to changes in organismal activity. These changes can also be made by genetic alterations. Theoretically, such perturbations are also reflected in the Metabolite CN and its analysis, which is not the case when relying solely on known pathways. In practice, Metabolite CN is beneficial in two key aspects. First, we can utilize it to uncover unknown information. Second, we build our known pathway pool in alignment with the Metabolite CN, avoiding the use of irrelevant pathways \cite{toubiana2013network}.

\subsection{Network Auralization}
\subsubsection{Network Auralization and Node Waveforms}
In 2022 Puzis \cite{puzis2023can} introduced an innovative application of sound recognition neural networks to predict centrality measures of nodes within a network, by learning from the ''sound'' emitted by network nodes. Networks can be likened to resonating chambers, where sound propagates through nodes and links, generating a waveform-based representation for every node. 

The core process of network auralization involves the propagation of energy among nodes to their neighbors until the total energy is evenly distributed throughout the entire network.

\subsubsection{M5 Architecture for Centrality Learning}
M5 is a very deep convolutional neural network, presented by Dai et al. \cite{dai2017Very} for recognizing environmental sounds in urban areas. M5 is built from four core blocks: $1X1$ Convolutional layer, Batch normalization and Max Pool layer. Average pooling layer and fully connected layer comes after the convolutional blocks, as presented in Supplementary Fig\ref{fig:M5}.  

Originally, M5 was designed for classification tasks with softmax as output \cite{dai2017Very}. For the purpose of his research, Puzis \cite{puzis2023can} transformed the M5 architecture into a prediction task architecture by replacing the softmax layer with a fully connected layer, and shaping it as a regressor. 

Puzis \cite{puzis2023can} trained several M5 models for various centrality measures using auralized random graphs, including Erdős-Rényi and Barabasi-Albert. Later, these models were tested on larger versions of the training graph types and also on well-known graphs like the Karate Club and the Southern Women Network. In the majority of cases, high correlations between the ground truth and predictions were observed, demonstrating that the combination of auralized nodes and M5 accurately reflect the network structure and the importance of nodes.

\section{Related Work}
\subsection{Predicting the Presence and Absence of Metabolic Pathways}
Predicting the presence and absence of metabolic pathways in an organism is a common approach in the attempt to learn an organism’s metabolic network. By leveraging a novel metabolic pathways classifier, we will be able to uncover pathways and bridge existing knowledge gaps in the metabolic network. 

Toubiana et al. \cite{toubiana2019combined} successfully classified metabolic pathways and identified undiscovered pathways in tomato pericarp. They constructed Metabolite CNs from three tomato datasets that were sampled across different seasons. Metabolite-level network features were extracted from these Metabolite CNs and aggregated into pathway-level features. Following feature selection, Machine Learning (\textbf{ML}) models were trained to detect known pathways of tomatoes, as well as pathways unrelated to tomatoes and random pathways. The Random Forest (\textbf{RF}) model yielded the best results, achieving an Area Under the Curve (\textbf{AUC}) score of 0.932 and an accuracy score of 83.78\%. The classification of candidate pathways identified 22 as potential tomato pathways. In vivo assays confirmed the presence of one of these pathways.

To the best of our knowledge and based on the available literature, Toubiana et al. \cite{toubiana2019combined} research stands as the only study to date that utilizes Metabolite CN structure to identify metabolic pathways. However, there is an extensive research effort dedicated to pathway classification. Dale et al. \cite{dale2010machine} were the pioneers in applying ML algorithms to detect metabolic pathways by learning from biological features, opening new options beyond the biological algorithms previously dominant in this field. Following the successful results, Dale highlighted the explainability of ML, which unlike alternative biological algorithms, enables researchers to understand why certain pathways were predicted to be present or absent. 

A decade after Dale’s publication, Basher et al. \cite{basher2020Metabolic} introduced the mlLGPR solution, which also apply ML to detect metabolic pathways, and enabled scalability in both the coverage of organisms and parallel computation. Other notable studies were made by Joe et al. \cite{joe2024Multi-label} and Aljarbou and Haron \cite{aljarbou2020determining}, which applied the XGBoost and Support Vector Machine models, respectively. Both studies achieved superior results compared to previous research.

In 2022 Shah et al. \cite{shah2022DeepRF} integrated a DL architecture with an ML classifying model. They fed Aljarbou and Haron’s \cite{aljarbou2020determining} features into a fully connected network with two hidden layers. The network learned the features and then used the network output to classify the presence or absence of pathways using RF model.

\subsection{Hyperlink Prediction}
One of the common use cases in hyperlink prediction methods’ evaluation is the classification of missing pathways in a metabolic network. 

\subsubsection{Hyperlink Prediction Taxonomy }
Chen and Liu \cite{chen2023survey} differentiate between “direct” methods that are specifically developed for hyperlink prediction and “Indirect” methods that were initially developed for other purposes such as traditional link prediction but applied to this task. They introduced a taxonomy of hyperlink prediction methods:

\noindent\textit{Similarity-based methods} 
rely on measuring the similarity between nodes, usually based on network features, and aggregate the similarity score of the hyperlink participants. Within this category, methods utilize metrics like Common Neighbors score and Katz Index score. 

\noindent\textit{Probability-based methods} often employ various probability estimators, such as maximum likelihood estimation and Bayes' theorem. Like Similarity-based methods, techniques in this category generate scores that often fail to capture and leverage high-order topological attributes. 

\noindent\textit{Matrix Optimization Methods} leverage incidence, adjacency or Laplacian matrices to design hyperlink prediction tasks as matrix optimization problems. Although this approach yields good results it suffers from significant disadvantages, notably its limited scalability and the inclusion of all candidate hyperlinks in the training.

\noindent\textit{DL methods} integrate the use of Node2Vec with widely used DL architectures such as GCN and GraphSage. DL methods have shown promising results in hyperlink prediction tasks by effectively capturing intricate relationships and structures within the network. 

Chen and Liu \cite{chen2023survey} assessed selected methods across multiple hypergraphs from diverse domains. DL methods dominated in terms of results compared to methods from other categories. Above all, a direct DL method called Chebyshev Spectral Hyperlink Predictor (\textbf{CHESHIRE}) \cite{chen2023Teasing} stood out by outperforming others in four out of five datasets, including a metabolic network dataset.

\section{Methods}
\subsection{NetAurHPD for Classification of Metabolic Pathways}
NetAurHPD combines graph-based and DL methods to predict the presence or absence of metabolic pathways in an organism based on its Metabolite CN. The full suggested pipeline is illustrated in Supplementary Fig\ref{fig:Augmentatiom process}. Inspired by Puzis \cite{puzis2023can} innovative work, we will develop a DL classifier based on aggregated metabolite signals, which function as hyperlinks. 

NetAurHPD requires two inputs, Metabolite CN (Supplementary Fig\ref{fig:NetAurHPD pipeline}.a) and a pool of labeled pathways (Supplementary Fig\ref{fig:NetAurHPD pipeline}.b), some of which occur in the organism and some do not. The Metabolite CN will be transformed into an aural representation (Supplementary Fig\ref{fig:NetAurHPD pipeline}.c) as described in Algorithm 1, resulting in a sound representation for each metabolite (Supplementary Fig\ref{fig:NetAurHPD pipeline}.d). Using the labeled pathways, we average the waveforms of each pathway's participants by index (Supplementary Fig\ref{fig:NetAurHPD pipeline}.e) and producing a single sound representation for each pathway (Supplementary Fig\ref{fig:NetAurHPD pipeline}.f). Based on these pathway sound representations, we will train M5 as classification model (Supplementary Fig\ref{fig:NetAurHPD pipeline}.g) to predict the class of each pathway (Supplementary Fig\ref{fig:NetAurHPD pipeline}.h).

The core of M5 has four convolution blocks with the components: convolutional layer, Relu activation function, Batch normalization, and Max Pool layer, as illustrated in Supplementary Fig\ref{fig:M5}. M5 learns each signal input through a sliding window technique. The signal propagates through the four convolutional blocks, followed by an average pooling layer and a fully connected layer, producing a single value. This value is then fed into a sigmoid activation function, yielding the final prediction.

\paragraph{Loss function} 
Binary Cross Entropy (\textbf{BCE}) is a common loss function in binary classification problems. BCE measures the performance of each computed output, by comparing the prediction value and the actual class, 0 or 1. The cross-entropy for each sample is calculated by the formula:
\vspace{-3pt}
\[
\text{BCE} = - \left( y_i \log(p(y_i)) + (1 - y_i) \log(1 - p(y_i)) \right)
\]

where \( y_i \) denotes the true class of the \( i \)-th sample. \( p(y_i) \) is the predicted probability of the \textit{Present} class for the \( i \)-th sample, and \( (1 - p(y_i)) \) is the probability of the \textit{Absent} class of the \( i \)-th sample. The formula measures how far the prediction is from the real class. The loss function is the mean of all samples’ BCE:
\vspace{-3pt}
\[
\text{BCELoss} = \frac{1}{N} \sum_{i=0}^{N} \text{BCE}
\]
\paragraph{Threshold optimization} 
In binary classification problem there is no guarantee that the commonly used threshold of 0.5 is the optimal cutoff point. Youden index is a statistical method used to learn the optimal threshold from the True Positive Rate (\textbf{TPR}) and False Positive Rate (\textbf{FPR}) results over different thresholds. The Youden index is calculated as:
\[
\text{Youden index} = \arg\max_x \{ \text{TPR} - \text{FPR} \}
\]
Based on the optimal threshold we will divide the prediction into classes and calculate the performance metrics \cite{liu2024joint}.

\subsection{Data Augmentation}
\subsubsection{Generation of Metabolite CN}
Frequently, DL models excel when trained on large datasets. However, training such a model on limited available data (low-sample-size regime) can be challenging. DL models trained over small datasets commonly struggle to generalize and may tend to overfit. Data augmentation methods aim to increase the size of the training data and face the low-sample-size regime challenge. Various approaches and methods exist for performing data augmentation. Some generate synthetic data using generative models or simulations, while others apply transformations to the existing data \cite{kebaili2023deep}. 

In our research, we imitate the metabolic activity within an organism, as illustrated in Supplementary Fig\ref{fig:Augmentatiom process}. This section describes the augmentation of the tomato pericarp’s metabolite profiles. Therefore, the augmentation process may differ for other organisms.

The required input for augmentation is the organism’s pathways in their directional format, allowing us to label the substrate and product metabolites in each pathway (Supplementary Fig\ref{fig:Augmentatiom process}.a). The next step involves creating different "Species" (Supplementary Fig\ref{fig:Augmentatiom process}.b), subgroups that represent organisms with variations in their biological activity due to environmental conditions or genetic manipulations \cite{toubiana2013network}. For each species, we generate several replicas, which are as the samples in the process (Supplementary Fig\ref{fig:Augmentatiom process}.c). Each sample is a closed metabolic system that starts with a fixed concentration level, uniformly distributed among the metabolites, meaning the number of units for each metabolite is not equal but similar. 

We mimic metabolic activity by executing a series of metabolic pathways (Supplementary Fig\ref{fig:Augmentatiom process}.d). In each iteration, if the participant metabolites are in sufficient amounts, substrate are transformed into products. The process can stop after a fixed number of iterations or when completing the pathways is impossible due to the lack of required metabolites. Once completed (Supplementary Fig\ref{fig:Augmentatiom process}.e) we calculate the Pearson Correlation Coefficient between every pair of metabolites (Supplementary Fig\ref{fig:Augmentatiom process}.f) to generate the Metabolite CN, referred to as Augmentations (Supplementary Fig\ref{fig:Augmentatiom process}.g). 
\subsubsection{Use Data Augmentation to Improve Model Learning}
As shown in the pipeline in Supplementary Fig\ref{fig:NetAurHPD pipeline}, the M5 model is trained on both the original Metabolite CN and the Augmentations (Supplementary Fig\ref{fig:NetAurHPD pipeline}.i). Similar to the original Metabolite CN, the auralized Augmentations (Supplementary Fig\ref{fig:NetAurHPD pipeline}.j) follow a pathways shape aggregation to produce pathway signals. Consequently, each pathway has several signals, with one derived from each original Metabolite CN or Augmentation (Supplementary Fig\ref{fig:NetAurHPD pipeline}.k). 

\section{Experiments}
\subsection{Dataset}
To study the suggested pathway prediction models we used the data provided by Toubiana et al. \cite{toubiana2019combined}. The architecture requires two inputs: 1- Metabolite CN constructed from organ samples, and 2– labeled pathways for the classification task.

\subsubsection{Metabolite CN construction} 
In our case, we used a dataset containing metabolic profiles of the tomato pericarp for three harvesting seasons: 2001,2003, and 2004. The data contains 4 to 6 biological replicates of every introgression line, which is a unique species developed by incorporating genes from one species into another, often from wild species into domestic crops. Each sample represents the concentration of a metabolite in the Introgression line. As a preliminary step to constructing the Metabolite CN, the data were normalized at both the sample and metabolite levels. Additionally, missing values were completed using a probabilistic PCA method available in the pcaMethods package in Rstudio. 

For every year a metabolite CN was constructed as described in section \ref{section:Metabolites CN}, where the selected correlation threshold was set to be $|r| \ge 0.3$ and $p \le 0.01$. For 2001 the Metabolite CN included 75 metabolites and 473 links. For 2003 it comprised 75 nodes and 869 edges, and for 2004 it had 78 nodes and 338 edges.

\subsubsection{Metabolic pathways for classification task} 
The dataset is divided into tomato and non-tomato pathways as positive and negative examples. Pathways were sourced from TomatoCyc, a pathway’s collection of tomatoes and from MetaCyc which includes non-plant pathways. Only pathways with more than two compounds that appear in the Metabolite CN were included in the dataset. Specifically, the three Metabolite CNs contain 52 common metabolites, and only pathways with two or more compounds from this list were added to the dataset. From TomatoCyc 169 pathways were identified as sharing two or more common metabolites. From MetaCyc 163 non-plant pathways were partially mapped to the common metabolites. To enhance model robustness, 85 random pathways were sampled as subsets of the 52 common metabolites and labeled as negative targets (negative sampling). It should be emphasized that we assume all TomatoCyc pathways are present in all three years.

\subsection{Training NetAurHPD on Tomato Metabolite CN}
\subsubsection{Experimental Setup - Experiment 1}
NetAurHPD will be tested separately for each year: 2001, 2003 and 2004, using its Metabolite CN. Additionally, in order to create a balanced target we had to down-sampling the negative class since there are 163 MetaCyc pathways, 85 random pathways and only 169 TomatoCyc pathways. For each training session, we randomly selected half of the negative examples from the MetaCyc pathways and the other half from the random pathways subset, resulting in a balanced positive-negative dataset.

Once we had the inputs, we followed NetAurHPD process to calculate the predictions (Supplementary Fig\ref{fig:NetAurHPD pipeline}). Given the relatively small size dataset, we trained M5 on all samples without the need to split it into batches. We evaluated the proposed solution with 10-fold cross-validation, where each pathway in the dataset appeared once in the validation set and received the model prediction. Based on all validation set predictions, we calculated the micro-average AUC and used Youden index to determine the optimal threshold and compute the performance metrics.

\paragraph{M5 Hyperparameter selection} 
This experiment involves a significant level of randomity. Negative examples are randomly chosen from MetaCyc and the random pathways for each new training session, and the 10-fold split of pathways is also based on random selection. Additionally, M5 weights and biases are initially set to random values. Furthermore, the use of Augmentations, which will be present later, includes an additional level of randomness. Due to these factors, each training session can vary significantly, making it impossible to compare models and optimizing parameters. Therefore, we used pre-defined parameters. The network dimensions are detailed in Supplementary Fig \ref{fig:M5}. We employed the Adam optimizer with a learning rate of 0.0001, a stride size of 8, and trained the model for 300 epochs. The experiments described later will follow the same experimental design, unless stated otherwise. 

\subsubsection{Experiment 1 Results}
\begin{figure}[h]
    \centering
    \includegraphics[width=1\linewidth]{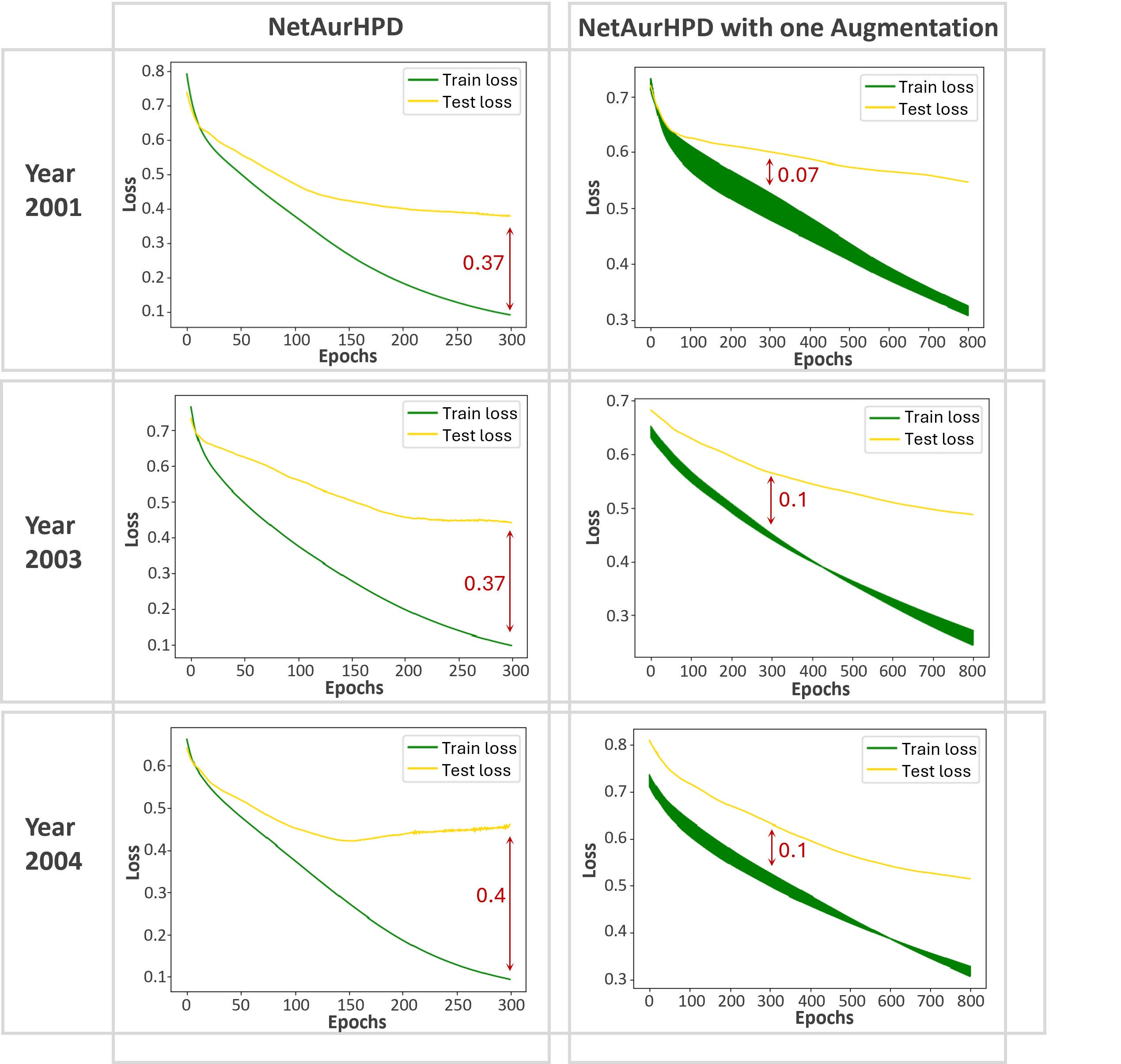}
    \caption{NetAurHPD and NetAurHPD with one Augmentation train and test loss charts of representing validation folds, across all years' datasets}
    \label{fig:loss}
\end{figure}

 The left column in Fig\ref{fig:loss} displays NetAurHPD loss charts of representing folds in the 10-fold cross-validation training. We can see that across the three years the model managed to learn and improve as the loss decreased. However, after approximately 150-200 epochs, the loss stabilized and then began to exhibit overfitting. In 2001 the micro-average AUC was 0.87 and the optimal threshold was set to 0.44, resulting in an accuracy of 0.80. For 2003 the micro-average AUC was 0.857 with an optimal threshold of 0.54 and an accuracy of 0.80, consistent with the 2001 results. In 2004 the micro-average AUC was 0.884 with an optimal threshold of 0.48 and accuracy of 0.813.

\paragraph{Results comparison to baseline} In general, NetAurHPD outperformed random model results and presented comparable results to Toubiana et al. \cite{toubiana2019combined}. Table \ref{table:NetAurHPD CV results} present metrics comparison by year between the abovementioned models. In 2001, Toubiana et al. performed better than NetAurHPD, demonstrating superior performance. In 2003, while Toubiana et al. \cite{toubiana2019combined} still had higher overall scores, NetAurHPD showed better precision and a lower FPR. Contrarily, in 2004, NetAurHPD outperformed Toubiana et al. \cite{toubiana2019combined} across all metrics.

\begin{table}[t]
\caption{Cross Validation Performance Metrics in Mild Conditions, by Model and Year}
\label{table:NetAurHPD CV results}
\begin{subtable}[t]{0.48\textwidth}
\scriptsize
\begin{tabular}{|p{2cm}|p{0.8cm}|p{0.8cm}|p{0.75cm}|p{0.5cm}|p{0.8cm}|p{0.8cm}|}
\hline
\rowcolor{gray!20} % Header row color
\textbf{\caption{2001\label{table:2001}}} & \textbf{Micro-Average AUC} & \textbf{Accuracy} &\textbf{TPR (Recall)} & \textbf{FPR} & \textbf{Precision} & \textbf{F1 Score} \\
\hline
\cellcolor{gray!20} CHESHIRE \cite{chen2023Teasing} & 0.9 & 0.84 & 0.786 & \textbf{0.1} & \textbf{0.88} & 0.83 \\
\hline
\cellcolor{gray!20} Toubiana et al. \cite{toubiana2019combined} &  \textbf{0.917} & \textbf{0.841}* & \textbf{0.841} & 0.159 & 0.841 & \textbf{0.841}  \\
\hline
\cellcolor{gray!20} NetAurHPD & 0.87 & 0.8 & 0.775 & 0.167 & 0.818 & 0.79 \\
\hline
\cellcolor{gray!20} NetAurHPD witn One Augmentation** & 0.87 & 0.8 & 0.8 & 0.182 & 0.814 & 0.8 \\
\hline
\end{tabular}
\end{subtable}
\begin{subtable}[t]{0.48\textwidth}
\scriptsize
\begin{tabular}{|p{2cm}|p{0.8cm}|p{0.8cm}|p{0.75cm}|p{0.5cm}|p{0.8cm}|p{0.8cm}|}
\hline
\rowcolor{gray!20} % Header row color
\textbf{\caption{2003\label{table:2003}}} & \textbf{Micro-Average AUC} & \textbf{Accuracy} &\textbf{TPR (Recall)} & \textbf{FPR} & \textbf{Precision} & \textbf{F1 Score} \\
\hline
\cellcolor{gray!20} CHESHIRE \cite{chen2023Teasing} &0.9 & \textbf{0.843} & 0.822 & 0.136 & \textbf{0.858} & \textbf{0.839} \\
\hline
\cellcolor{gray!20} Toubiana et al. \cite{toubiana2019combined} & \textbf{0.91} & 0.823* & \textbf{0.823} & 0.177 & 0.827 & 0.823  \\
\hline
\cellcolor{gray!20} NetAurHPD & 0.857 & 0.8 & 0.745 & 0.136 & 0.845 & 0.79 \\
\hline
\cellcolor{gray!20} NetAurHPD witn One Augmentation** & 0.852 & 0.789 & 0.714 & \textbf{0.134} & 0.841 & 0.77 \\
\hline
\end{tabular}
\end{subtable}
\begin{subtable}[t]{0.48\textwidth}
\scriptsize
\begin{tabular}{|p{2cm}|p{0.8cm}|p{0.8cm}|p{0.75cm}|p{0.5cm}|p{0.8cm}|p{0.8cm}|}
\hline
\rowcolor{gray!20} % Header row color
\textbf{\caption{2004\label{table:2004}}} & \textbf{Micro-Average AUC} & \textbf{Accuracy} &\textbf{TPR (Recall)} & \textbf{FPR} & \textbf{Precision} & \textbf{F1 Score} \\
\hline
\cellcolor{gray!20} CHESHIRE \cite{chen2023Teasing} & \textbf{0.89} & \textbf{0.822} & 0.816 & \textbf{0.171} & \textbf{0.826} &\textbf{ 0.821} \\
\hline
\cellcolor{gray!20} Toubiana et al. \cite{toubiana2019combined} & 0.876 & 0.761* & 0.761 & 0.239 & 0.766 & 0.76  \\
\hline
\cellcolor{gray!20} NetAurHPD & 0.884 & 0.813 & \textbf{0.834} & 0.2 & 0.8 & 0.817 \\
\hline
\cellcolor{gray!20} NetAurHPD witn One Augmentation** & 0.86 & 0.797 & 0.825 & 0.23 & 0.781 & 0.8 \\
\hline
\end{tabular}
\begin{tablenotes}
    \item * The accuracy was not directly presented in the article but was deduced from the information provided. 
    \item ** Trained over 126 pathways, as described in subsection \ref{subsubsection Experimental Setup - Experiment 2}.
\end{tablenotes}
\end{subtable}
\end{table}

\subsection{Use Data Augmentation During Training}
DL models often require large amounts of data to reach high performance and prevent overfitting. The process of constructing Metabolite CNs, starting from assembling plant populations, is complex, time-consuming, and sometimes even infeasible. Additionally, other organisms might have fewer known pathways and smaller labeled datasets. For these reasons, we aim to generate more data and enlarge the training set.

\subsubsection{Experimental Setup - Experiment 2}
\label{subsubsection Experimental Setup - Experiment 2}
The augmentation process required pathways known to exist in the organism in their directional format, allowing us to identify the substrates and products. We created a directional pathway pool of 126 pathways out of the 169 known pathways of tomato pericarp. Based on these pathways, we generate 10 Augmentations to be used in the training process.

The design of the current experiment is based on the setup of experiment 1, however here we trained five different models for each validation fold. The first model was trained on pathways’ signals from the original Metabolite CN. The other four models were trained also over signals from the Augmentations. As described above, in each epoch the model learned the signals from every Metabolite CN. The four models differed in the number of Augmentations included in the training: 1, 3, 5, and 7. For every trained model, the Augmentations were randomly chosen from the 10 pre-generated Augmentations.

We conducted 20 sessions of 10-fold cross-validation, each including the modeling of the 5 different models. Hyperparameters in this experiment are the same as in experiment 1 except that the model learned over 800 epochs with a learning rate of 0.00001 to enable smoother convergence for the larger models. 

As mentioned, a basic assumption in our experiment is that all positive pathways exist in the organism from which the Metabolite CN is derived. Therefore, for this experiment, we used only 126 positive examples and randomly chose 126 negative examples, as done in experiment 1. Since 43 known tomato pathways were dropped from the trainset, some metabolites did not appear in the Augmentations and yet were involved in negative pathways. Hence, we were missing the waveforms for those metabolites and could not aggregate them into the pathway representation. To address this issue, we manually added the missing metabolite to every sample (Supplementary Fig\ref{fig:Augmentatiom process}.c). Ultimately, all of them had a degree of 0 in the final Augmentation and had a waveform of ones, being not connected to any node and unable to propagate the impulse.   

\subsubsection{Experiment 2 Results}
The right column of Fig\ref{fig:loss} displays NetAu-
rHPD with one Augmentation loss charts representing folds from this experiment. We observe that the learning of NetAurHPD with Augmentation is prolonged compared to NetAurHPD alone, with a lower level of overfitting. 

\begin{figure}[t]
    \centering
    \includegraphics[width=1\linewidth]{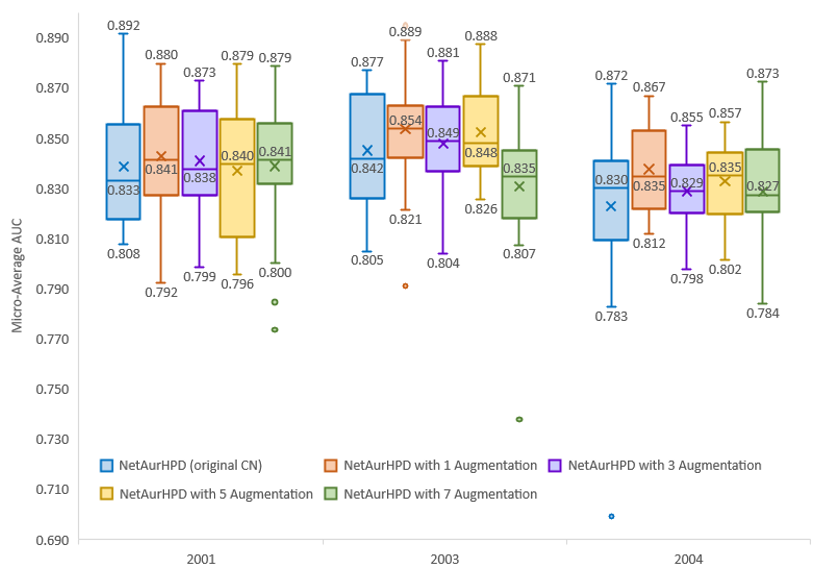}
    \caption{Micro-Average AUC of NetAurHPD sessions with Augmentations, by Model and Year}
    \label{fig:augmentations results}
\end{figure}

As presented in Fig\ref{fig:augmentations results}, almost all groups of models trained with 1,3 or 5 Augmentations presented equal or higher median micro-average AUC compared to NetAurHPD. For the three years, the NetAurHPD with Augmentations appears more robust and stable, with smaller interquartile ranges compared to NetAurHPD alone. Specifically, the improvement is noticeable in the second quartile. In this experiment, NetAurHPD with one Augmentation (orange group in Fig\ref{fig:augmentations results} presented better performances considering the median, mean, maximum AUC and box-and-whiskers boundaries. Therefore, there is no justification to use NetAurHPD with more than one Augmentation.

Additionally, the chart reveals how the randomness of the model and experiment affect the models’ quality, with range between higher and lower whiskers varying from 0.055 to 0.089. However, for NetAurHPD with one Augmentation, the range decreases from an average of 0.082 for NetAurHPD alone to an average of 0.07.

\subsection{NetAurHPD Advantages - Prediction in Tough Realistic Conditions}
In real-world conditions, biologists may face organisms with a limited number of known metabolic pathways \cite{toubiana2020correlation}. Identifying unknown pathways in these cases is significantly more challenging than in well-explored organisms, where hyperlink prediction methods relying solely on known pathways can be more effectively applied. However, the Metabolite CN of such organisms can provide valuable additional information, which NetAurHPD can leverage to enhance prediction outcomes.
\subsubsection{Experimental Setup - Experiment 3}
In this experiment, we examine scenarios where only a limited number of pathways are known, and there are metabolites that do not appear in any pathway. We compare the performance of state-of-the-art CHESHIRE model and NetAurHPD in two versions: one using the original Metabolite CN and another using also one Augmentation. While NetAurHPD uses both Metabolite CN and pathways as input, CHESHIRE relies solely on pathways.

From the tomato pericarp dataset, we selected 32 metabolites involved in a minimal number of pathways. We excluded all pathways that include these metabolites from the training set and created a test set consisting of 40 TomatoCyc pathways, 27 MetaCyc pathways, and 13 random pathways. For each year, we conducted 10 sessions, where in each session we trained the models on a larger portion of the training set, starting at 10\% and increasing to 100\%. The complete training set includes 129 TomatoCyc pathways, 65 MetaCyc pathways, and 64 random pathways. In every training session, all models were trained on the same dataset and the classification threshold set to 0.5. NetAurHPD models were trained over 100 epochs and CHESHIRE over 75 epochs. The other parameters for CHESHIRE were configured according to the recommendations provided in the full source code included with the article. 

\subsubsection{Experiment 3 Results}
\begin{figure}[t]
\begin{subfigure}[t]{0.48\textwidth}
    \centering
    \includegraphics[width=1\linewidth]{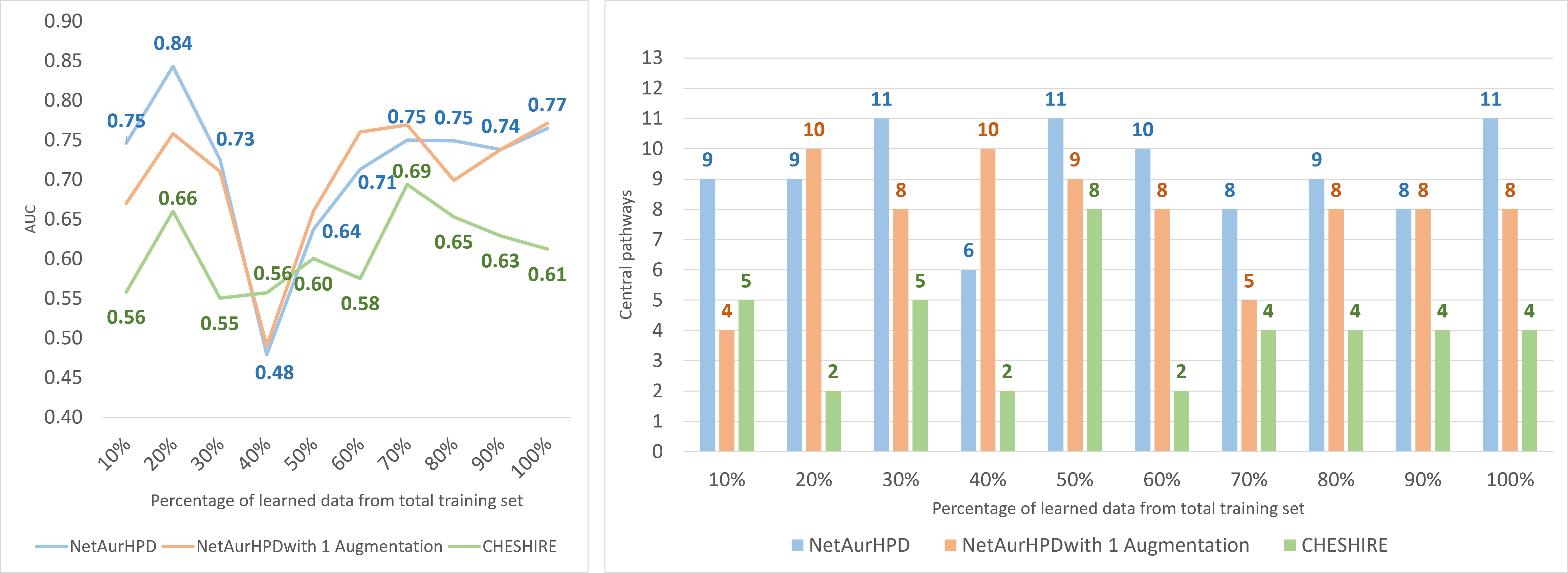}
    \caption{2001}
    \label{fig:Tough Realistic Condition 2001}
\end{subfigure}
\begin{subfigure}[t]{0.48\textwidth}
    \centering
    \includegraphics[width=1\linewidth]{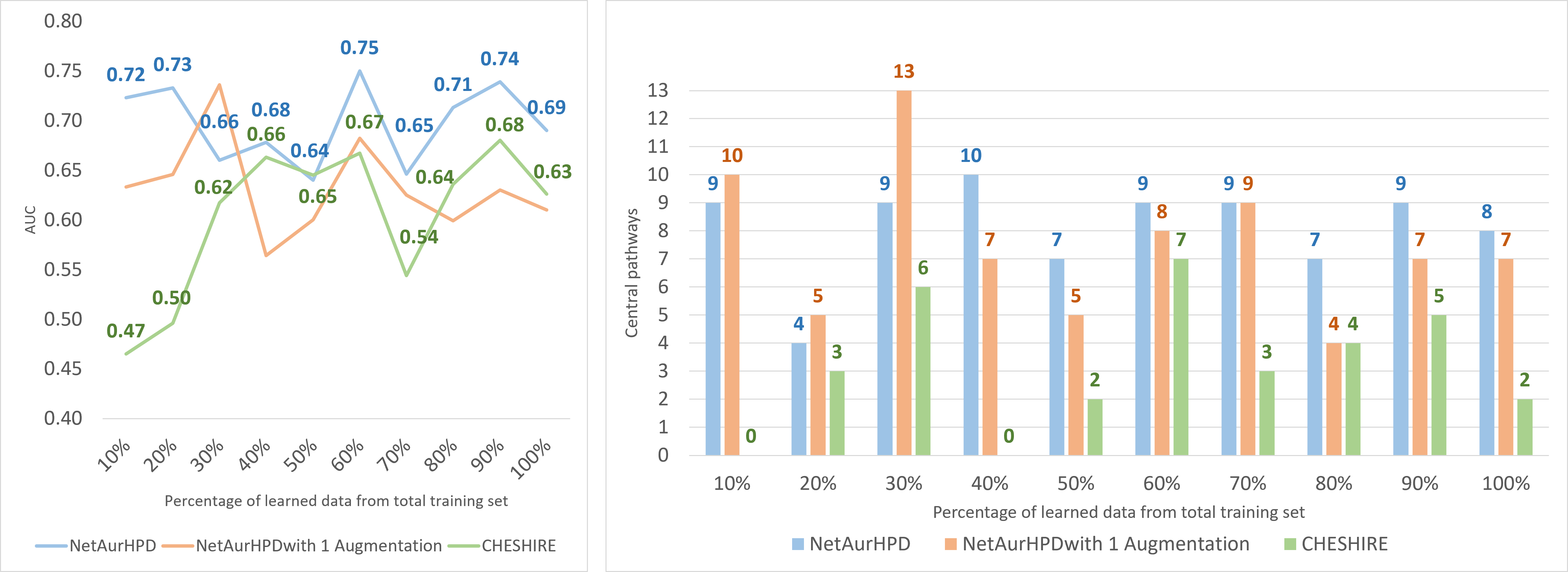}
    \caption{2003}
    \label{fig:Tough Realistic Condition 2003}
\end{subfigure}
\begin{subfigure}[t]{0.48\textwidth}
    \centering
    \includegraphics[width=1\linewidth]{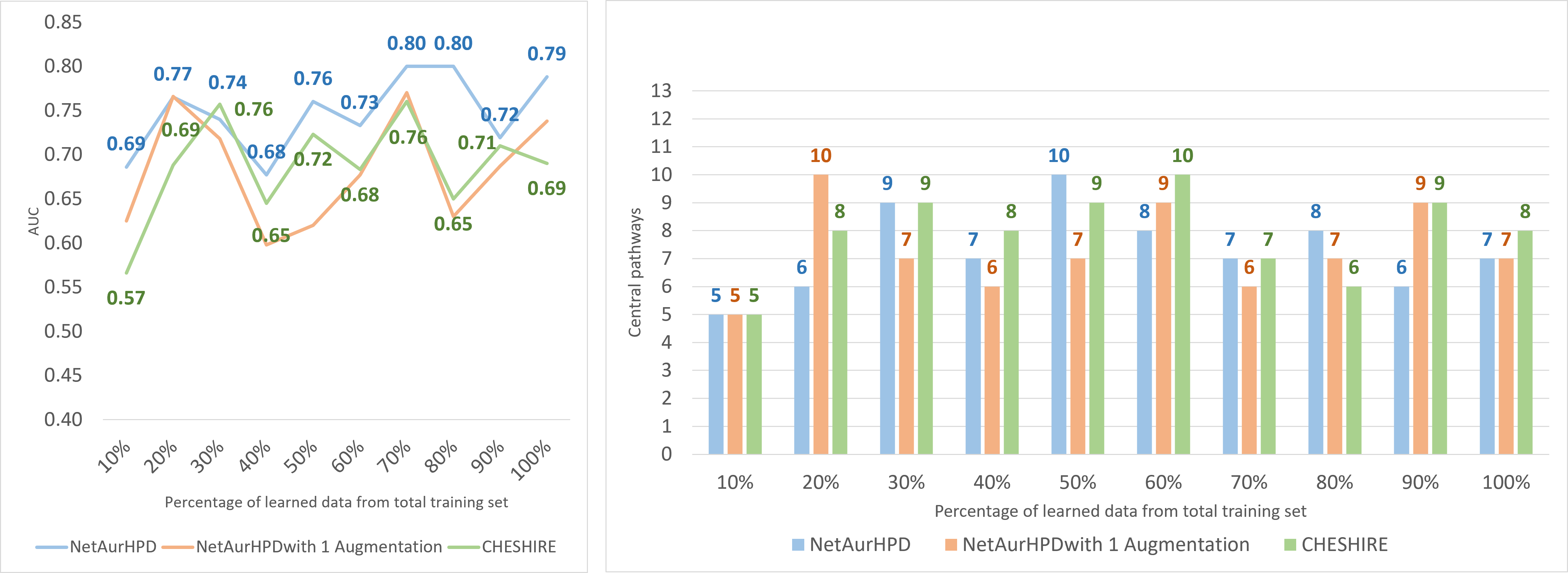}
    \caption{2004}
    \label{fig:Tough Realistic Condition 2004}
\end{subfigure}
\caption{Results comparison of CHESHIRE vs. NetAurHPD with and without augmentation on tough realistic conditions.
Left column figures: AUC Comparison, by Model.
Right column Figure: True Classification of Central Pathways, by Model.
}
\end{figure}

The left side of Figures \ref{fig:Tough Realistic Condition 2001},\ref{fig:Tough Realistic Condition 2003} and \ref{fig:Tough Realistic Condition 2004} shows the AUC performance of each model, with the x-axis representing the percentage of actual data used for training from the total train set and the y-axis representing the AUC. We can see that NetAurHPD requires a smaller amount of pathways to achieve satisfactory AUC, while CHESHIRE struggles to reach high scores in many training sessions. Generally, in all three years NetAurHPD outperforms CHESHIRE. 
As described above the test set includes 40 TomatoCyc pathways, 13 of which are identified as central pathways critical for the organism's survival. In addition to the AUC score analysis, we compared the number of true predictions for these central pathways. The results, presented on the right side of Figures \ref{fig:Tough Realistic Condition 2001},\ref{fig:Tough Realistic Condition 2003} and \ref{fig:Tough Realistic Condition 2004}, show that in 2001 and 2003 both NetAurHPD models correctly predicted a high number of central pathways, while CHESHIRE produced poor results. In 2004, CHESHIRE improved and achieved competitive results in many sessions.

\subsection{Discussion}
As shown in Fig\ref{fig:loss}, NetAurHPD successfully learned the waveforms of positive and negative pathways. Moreover, the use of Augmentations improved the learning process and prevented overfitting. 
Compared to the baseline, as detailed in Tables \ref{table:2001} - \ref{table:2004}, NetAurHPD models achieved comparable results to those of Toubiana et al. \cite{toubiana2019combined}. In 2001 Toubiana et al. outperformed NetAurHPD models in all metrics. However, in 2003, NetAurHPD models achieved better precision and FPR scores, indicating a more conservative model. From a broader perspective, precision is particularly significant because the following step after prediction is to conduct in vivo validation. High precision is crucial to minimize experiments on false positive pathways, thus saving time and resources. For 2004, NetAurHPD surpassed the baseline across all metrics.

As shown in Fig\ref{fig:augmentations results}, the use of Augmentation enhances the models' performance and robustness. Not only is the variance lower, but the median micro-average AUC is also higher.

Experiment 3 demonstrated the advantage of using NetAurHPD with Metabolite CN as input. As presented in Tables \ref{table:2001} - \ref{table:2004}, CHESHIRE \cite{chen2023Teasing} yielded outstanding results in mild conditions. However, in tough realistic conditions as simulated in experiment 3, CHESHIRE's performance significantly declined. Under those conditions, the Metabolite CN plays a central role in learning. Using only the pathways as input, CHESHIRE struggles to learn pathways involving metabolites which are not included in its sole input. In contrast, NetAurHPD requires a low number of pathways to achieve a sufficient score due to the structural information embodied the signals. 

More broadly, Figures \ref{fig:Tough Realistic Condition 2001},\ref{fig:Tough Realistic Condition 2003} and \ref{fig:Tough Realistic Condition 2004} highlight the challenging task of pathway prediction, as the models do not consistently improve with more data. In some sessions, the AUCs of all models decreased, demonstrating the complexity of the problem.

Analyzing the central pathways, it is surprising that "rare" metabolites are part of such critical pathways. This underscores the importance of gaining information about all known metabolites. As shown in Figures \ref{fig:Tough Realistic Condition 2001},\ref{fig:Tough Realistic Condition 2003} and \ref{fig:Tough Realistic Condition 2004}, both NetAurHPD models outperformed CHESHIRE in most of the sessions, even though NetAurHPD with one Augmentation presented low AUC.

\section{Conclusions}
In this research, we developed NetAurHPD method to predict the presence or absence of metabolic pathways in an organism based on its Metabolite CN. We trained NetAurHPD on three datasets of tomato pericarp’s Metabolite CN and achieved comparable results to the expert-based feature extraction model presented by Toubiana et al. \cite{toubiana2019combined}.    

Additionally, we aimed to develop a data augmentation method to enlarge the training dataset, which is often relatively small for many organisms. Data augmentation is particularly crucial for DL models, which generally require large datasets to achieve optimal performance. In this research, we augmented the dataset by imitating the organism’s metabolic activity to generate additional  Metabolite CNs. This augmentation technique improved NetAurHPD's performance under mild conditions, prolonged its learning abilities, and mitigated overfitting.

According to our results, the advantage of NetAurHPD is in challenging real-world conditions where biologists are exploring organisms with limited prior knowledge of known pathways. In such cases, NetAurHPD outperforms the state-of-the-art hyperlink prediction method CHESHIRE, despite CHESHIRE's outstanding results under mild conditions. The results demonstrate the ability of the Network Auralization method to learn the graph structure not only at the node level but also at the subset level. Additionally, the results highlight the significance of using Metabolite CN as a model input alongside pathway examples.

However, NetAurHPD has a low level of explainability and cannot provide biologists with a full understanding of its results. 
Alternatively, Toubiana et al. model  \cite{toubiana2019combined} offers a higher level of explainability due to its metabolite-level network features. In future work, NetAurHPD should be examined for its sensitivity to organismal perturbations, which are theoretically reflected in the Metabolite CN, by evaluating the model’s ability to recognize changes in organismal activity. Further investigation should also explore the potential of using NetAurHPD for transfer learning.

\bibliography{thesis}

%\begin{appendices}
% Manually format the URL
\appendix
\section{Code availability}
NetAurHPD source code used train hyperlink prediction classifier was deposited on GitHub: \url{https://github.com/TamirBar-Tov/NetAurHPD-Network-Auralization-Hyperlink-Prediction-Method}.

\section{Methods}
\subsection{Network Auralization Algorithm}
\label{appendix: Network Auralization Algorithm}
In formal terms, $\smash{(G = (V, E)}$) denotes an undirected and unweighted graph, where \(V\) represents the set of nodes and \(E\) the set of edges. Let \(s_{v,t} \in \mathbb{R}\) denote the amount of energy held by node \(v \in V\) at time \(t\), considered as the node's potential. The vector $\smash{\mathbf{S}_t = (s_v : v \in V)}$ represents the vector of potentials of all nodes at time \(t\). In every iteration of the diffusion process, each node potentially exchanges energy with its neighbors. The delta between the quantities determines the updated potential at time \(t\) compared to time \(t-1\).

Let \(A\) stand for the adjacency matrix of \(G\), and \(D = \sum_{v}A{\scriptstyle v}\) denote the vector of nodes degree. Define \(P_{u,v} = {A_{u,v}}/{D_u}\) to be the power that \(u\) may apply on \(v\), which remains constant over time (unlike the potential). The amount of energy that node \(u\) applies to every neighbor \(v\) at time \(t\) is: $\Delta S_{t,u,v} = S_{t-1,u} \cdot P_{u,v}$. In matrix form, \(Diag(\mathbf{S}_t)\) is an \(n \times n\) matrix with values of \(\mathbf{S}_t\) along the diagonal and \(\Delta S_t = \text{Diag}(\mathbf{S}_{t-1}) \times P\). The potential vector is updated as:
\vspace{-4pt}
\[
\mathbf{S}_t = \mathbf{S}_{t-1} + \left( \sum_{u \in V} \Delta S_{t,u,v} - \sum_{u \in V} \Delta S_{t,v,u} \right),
\]
where \(\sum_{u \in V} \Delta S_{t,u,v}\) is the incoming energy to node \(v\) and \\ \(\sum_{u \in V} \Delta S_{t,v,u}\) is the outgoing energy flow. Moreover, the stabilization process is extended by preserving a fraction \(m\) of the energy from the previous iteration. The energy flow will now be represented as: $\Delta S_{t,u,v} = S_{t-1,u} \cdot P_{u,v} + m \cdot \Delta S_{t-1,u,v}$.

Algorithm 1 presents the pseudo-code of network auralization adapted for PyTorch implementation, the full source code is available on GitHub: \url{https://github.com/puzis/centrality-learning} (accessed on 23 March 2023). The operator $T$ is matrix transpose. The operators $sum$ and $mean$ aggregate elements of a matrix along the dimension specified by $dim$. For our research, the DC component which refers to the values on which $S$ stabilizes after impulse response, is not of interest and will be removed from the vector. 

\begin{figure}[h]
    \centering
    \includegraphics[width=1\linewidth]{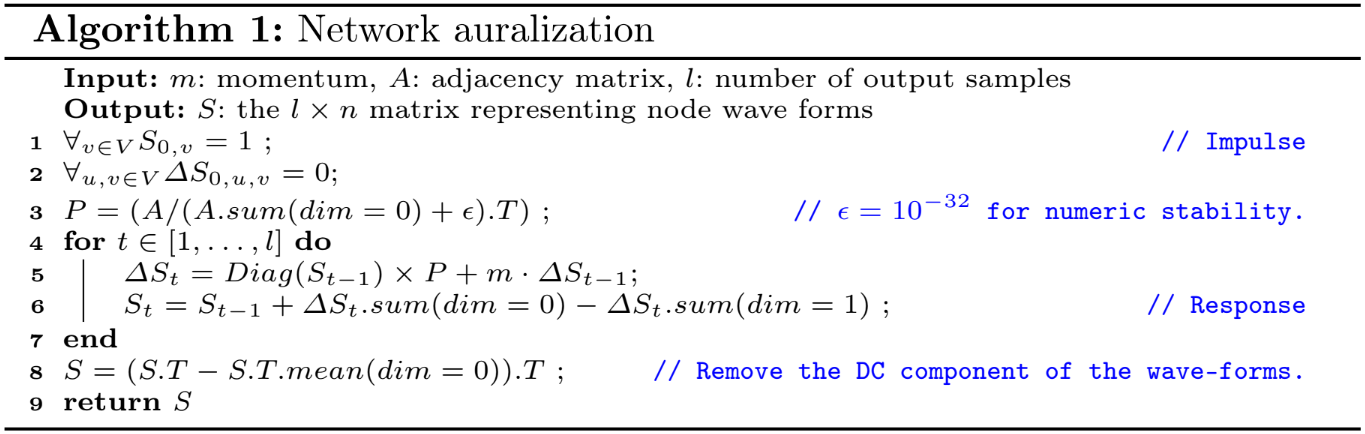}
    \label{fig:algorithm_1}
\end{figure}

\subsection{Performance Metrics and Models Comparison} 
Given a set of probability predictions, the classes are often determined based on a defined threshold. A variety of performance metrics can be used to evaluate the predicted classes and later compare models. True Positive Rate (\textbf{TPR}), also known as recall, measures the ratio between the true positive predictions and all actual positive samples. False Positive Rate (\textbf{FPR}) measures the ratio between the false positive predictions and all actual negative samples. Accuracy represents the proportion of correct predictions out of all samples \cite{fawcett2006roc}. 

Moreover, another common criterion for comparing models is based on the receiver operating characteristic (\textbf{ROC}) curve, which does not depend on a predefined threshold. ROC curve is created by plotting the TPR on the y-axis as a function of the FPR at different threshold values. The AUC indicates the model's performance level, where an AUC score of 1 means a perfect classifier and an AUC score of 0.5 indicates that the classifier is no better than a random classifier (for a balanced target).

\section{Supplementary Figures}
%\subsection{NetAurHPD Pipeline}

\begin{figure}[H]
    \centering
    \includegraphics[width=5cm]{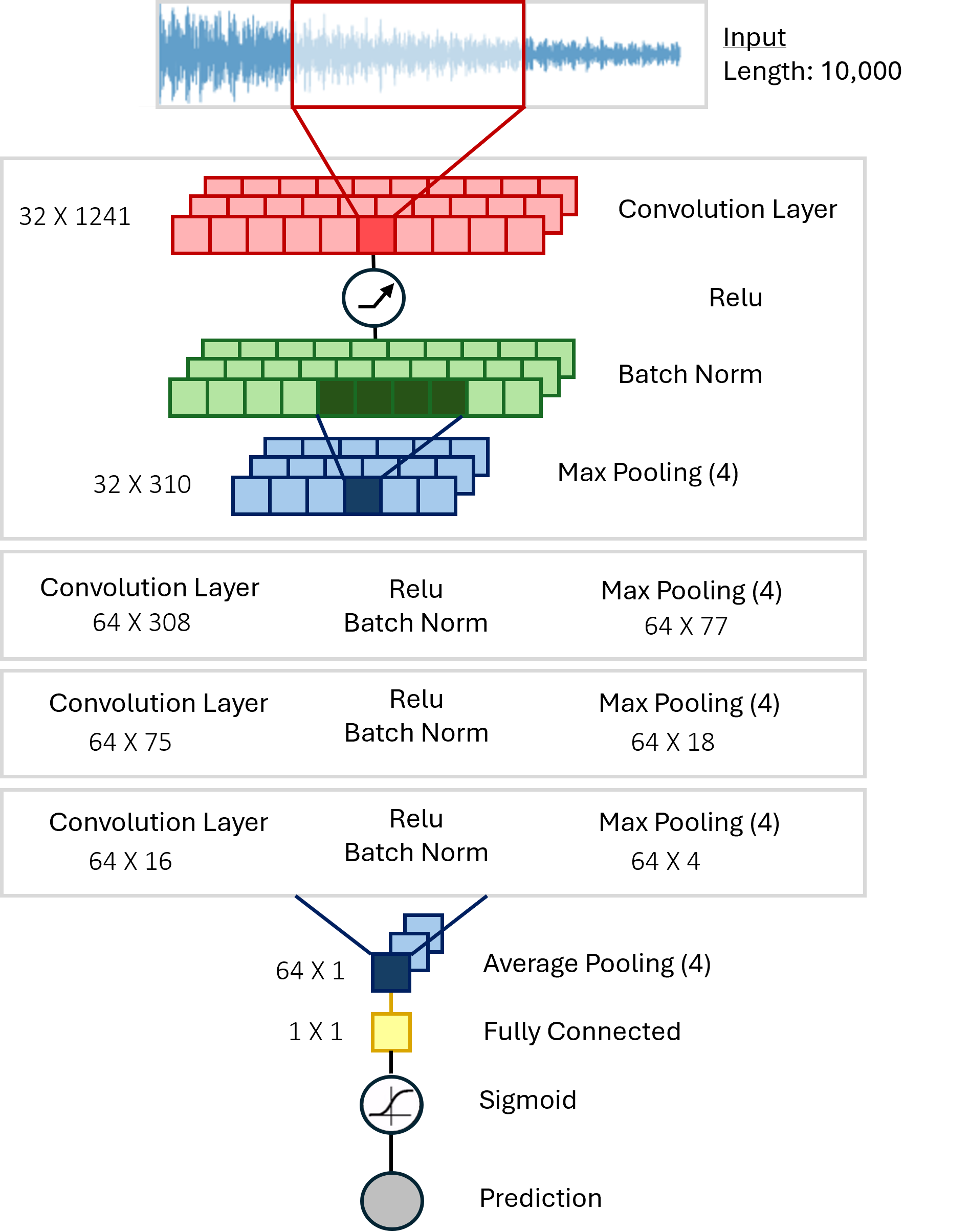}
    \caption{\label{fig:M5}M5 Deep Convolutional Neural Network Architecture for Pathways Classification}
\end{figure}

\begin{figure*}[h]
    \centering
    \includegraphics[width=0.9\textwidth]{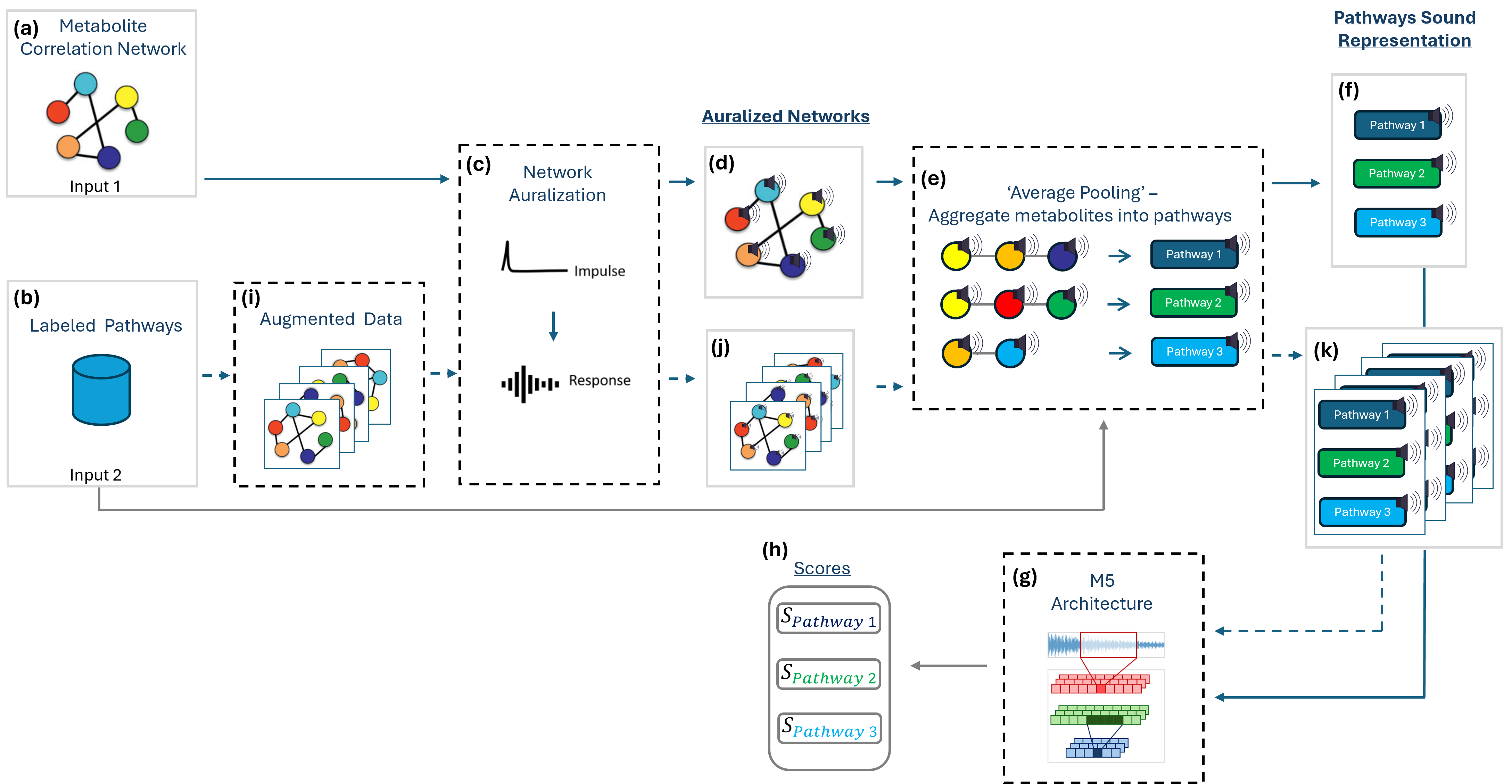}
    \caption{\label{fig:NetAurHPD pipeline}NetAurHPD full pipeline to predict the presence or absence of metabolic pathways in an organism based on its Metabolite CN. (a) First Input of Metabolite CN. (b) Second input of labeled pathways. (c) Network Auralization. (d) Waveforms for metabolite. (e) Average aggregation which transforms the sinal from metabolite level to pathways level. (f) Waveforms for each pathway. (g) M5 algorithm. (h) Final output. (i) Generation of Augmentations. (j) Waveformsn for metabolite from the Augmentations. (k) Waveforms for each pathway from the Augmentations.}
\end{figure*}

%\subsection{Data Augmentation Full Pipeline}

\begin{figure*}[h]
    \centering
    \includegraphics[width=0.70\textwidth]{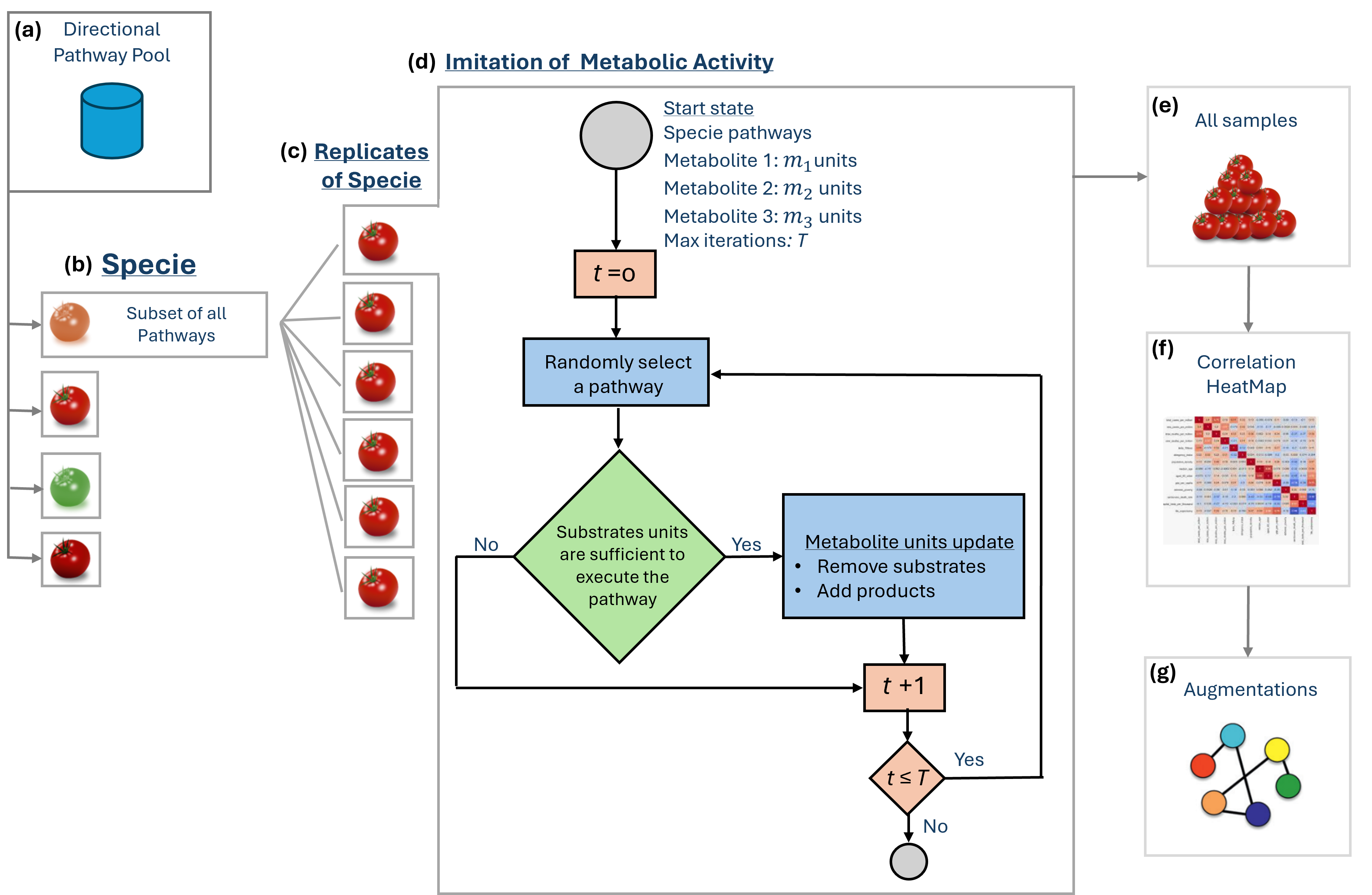}
    \caption{\label{fig:Augmentatiom_process}Data Augmentation workflow to generate Metabolite CN. (a) Collection of directional pathways. (b) Creation of Species, which represent the variants of the organism in nature. (c) Generate a few samples of every species. (d) Imitate the metabolic activity in an organism, starting with a fixed concentration level. In every iteration execute one reaction and update the substrates and products concentration. (e) The final state of the samples is reached when the process is stopped after a predefined number of iterations or when the execution of reactions is no longer feasible. (f) Calculate Pearson Correlation Coefficient between each pair of metabolites. (g) Construct Augmentations.}
    
\end{figure*}

%\end{appendices}

\end{document}